\documentclass{aa}
\usepackage{latexsym}
\usepackage{graphicx}
\usepackage{natbib}

\newcounter{IonCS}

\newcommand{\eqn}[1]{(\ref{#1})}
\newcommand{\fig}[1]{Fig.\,\ref{#1}}
\newcommand{\tab}[1]{Table\,\ref{#1}}
\newcommand{\sect}[1]{Sect.\,\ref{#1}}

\def\figwid{88mm}
\def\figwff{95mm}

\newcommand{\NEW}[1]{{{#1}}}

\begin{document}


\title{\mbox{Link between the chromospheric network}
       \mbox{and magnetic structures of the corona}} 

\titlerunning{Link between chromospheric network and corona}
\authorrunning{Jendersie \& Peter}

\author{Stefan Jendersie \& Hardi Peter}
\offprints{H.~Peter}

\institute{Kiepenheuer-Institut f\"ur Sonnenphysik,
                79104 Freiburg, Germany; {\tt peter@kis.uni-freiburg.de}}

\date{Received 2.\,Aug.\,2006 / Accepted 11.\,Sept.\,2006}

\abstract%
%
{%
Recent work suggested that the traditional picture of the corona
above the quiet Sun being rooted in the magnetic concentrations of the
chromospheric network alone is strongly questionable.
}
{%
\NEW{Building on that previous study we} explore the impact of magnetic
configurations in the photosphere and the low corona on the magnetic
connectivity from the network to the corona.
\NEW{Observational studies of this connectivity are often
  utilizing magnetic field extrapolations.
  However, it is open to which extent such extrapolations really
  represent the connectivity found on the Sun, as observations are
  not able to resolve all fine scale magnetic structures.
  The present numerical experiments aim at contributing to this question.}
}
{%
We investigated random salt-and-pepper-type distributions of
kilo-Gauss internetwork flux elements carrying some $10^{15}$ to
$10^{17}$\,Mx, which are hardly distinguishable by current
observational techniques.
\NEW{These photospheric distributions are then extrapolated into the
  corona using different sets of boundary
  conditions at the bottom \emph{and} the top.
This allows us to investigate} the fraction of network flux which is
connected to the corona, as well as the locations of those
coronal regions which are connected to the network patches.
}
{%
%
%
 \NEW{We find} that with current instrumentation one cannot
really determine from observations, which regions on the
quiet Sun surface, i.e.\ in the network and internetwork, are
connected to which parts of the corona through extrapolation
techniques.
\NEW{Future spectro-polarimetric instruments, such as with Solar\,B
  or {\sc{Gregor}}, will provide a higher sensitivity, and studies
  like the present one could help to estimate to which extent one
  can then pinpoint the connection from the chromosphere to the corona.}
}
{%
}
%
\keywords{     Sun: magnetic fields
           --- Sun: atmosphere
           --- Sun: corona} 

\maketitle

\section{Introduction}\label{intro}

The appearance of the large-scale magnetic field on the quiet Sun
photosphere is dominated by the so called network structure, which is
found at the boundaries of the super-granular cells.
This magnetic network with a typical scale of 20\,Mm is made
up by concentrations each carrying a magnetic flux in the range of
$10^{18}$ to $10^{19}$\,Mx with field strengths of the order of
kilo-Gauss \citep{Schrijver+al:1997sp}.
The region in-between these network structures, the internetwork, is
not field-free \citep{Livingston+Harvey:1975}, but shows a small
average flux density appearing to be of the order of a few to
50\,Mx/cm$^2$.
Depending on the spatial resolution, polarimetric sensitivity and
diagnostic technique (Hanle and Zeeman effect), authors find a weak
volume filling component \citep{Faurobert+al:2001}, or small
concentrations with kilo-Gauss flux tubes
\citep[e.g.][]{Dominguez-Cerdena+al:2003,Sanchez-Almeida+al:2003}.

As there are flux tubes with high, kilo-Gauss field strengths to
be found in the network as well as the internetwork, a better concept
to distinguish between the network and internetwork would be the
amount of flux carried by each individual flux elements.
With some $10^{16}$ to $10^{17}$\,Mx each internetwork (flux tube)
element carries  about one to two orders of magnitude less magnetic
flux than the network patches
\citep{Lin+Rimmele:1999,Socas-Navarro+Sanchez-Almeida:2002}.

Even though each internetwork patch carries a very small amount of
flux as compared to the stronger network patches, the weak
internetwork patches can contribute a significant amount to the
total (unsigned) flux on the solar surface, because they are so
abundant.
In contrast to the old canopy concept
\citep[e.g.][]{Giovanelli:1980}, where the internetwork patches have
been neglected, the abundant low flux but high field strength
concentrations in the internetwork do interact with the strong
network patches: a sizable fraction of the field lines originating
from the internetwork patches do not close back to another
internetwork patch very close by, but can connect either to the
stronger network patches or to the corona.

This was first pointed out by \cite{Schrijver+Title:2003} who
constructed a network patch surrounded by many smaller internetwork
patches by placing magnetic charges at a plane, performed a potential
field extrapolation and then studied the connections from their
simulated photosphere, i.e.\ the distribution of point charges, to
the corona.
They found that, depending on the average absolute flux density, only
part of the network flux is actually connected to the corona.
At an internetwork average absolute flux density of about
20\,Mx/cm$^2$, which is compatible with solar observations, they
found that only half of the network flux connects into the corona,
the other half connects to the surrounding internetwork.
Consequently, half of the magnetic flux in the corona originates from
the internetwork, which is in strong contrast to the traditional canopy
picture.

In this paper we will go one step further than
\cite{Schrijver+Title:2003} and not only demonstrate that the coronal
magnetic field to a large part originates in the internetwork, but
also investigate the role of the (spatial) distribution of the
magnetic field in the low corona and the photosphere, i.e.\ at the
upper and lower boundary, for the link between the photosphere and
the corona.
This investigation will show the limitations for the knowledge we might
gain from magnetic field extrapolations on the connections from the
network and internetwork to the corona.
As it will turn out, this is basically through our imperfect ability
to determine the photospheric magnetic field at small enough scales
with sufficient polarimetric sensitivity.

The organization of this paper is as follows.
In order to also employ a boundary condition at the top of the
computational domain we briefly describe a Fourier transform technique
for the  potential field extrapolation, the boundary conditions
representing the photosphere and the low corona as well as how to
derive the fraction of network flux reaching the corona
(\sect{method}).
Based on these results for the extrapolations with different boundary
conditions we first compare our results to previous work in
\sect{comparison} before we discuss our new findings on the impact of the
low corona (\sect{impact_ubc}) and the distribution of photospheric
flux (\sect{impact_ubc}) on the magnetic connection from the
photosphere to the corona.
We will conclude the paper with a discussion of the consequences of
this study for further investigations of the chromosphere--corona
coupling in \sect{conclusions}.

\section{Magnetic field extrapolation and boundary conditions} \label{method}

In order to determine the magnetic coupling between internetwork fields
and the lower corona we will calculate the fraction of the magnetic
flux originating from the internetwork photosphere reaching into the
corona.
In a previous study \cite{Schrijver+Title:2003} computed ``field lines
by tracing the path of monopolar test particles through the summed potential
field above a plane [resembling the photosphere] filled with
magnetic point charges of mixed polarities that surround a small
number of significantly stronger point charges'' establishing a network like
configuration.
To enforce a vertical field at a coronal height, they added mirror charges at 
a height of $z{=}100$\,Mm.
The goal of the present paper is to re-visit the problem posed by
\cite{Schrijver+Title:2003} in order to further  investigate the
impact of the boundary conditions at the top, i.e. in the corona, and
at the bottom, i.e. in the photosphere.
Thus we will apply a magnetic field extrapolation allowing the
allocation of reasonable boundary conditions at the bottom {\it and}
the top of the computational domain.
Within the framework of the study of \cite{Schrijver+Title:2003} this
would not be possible for the upper boundary.
For the sake of simplicity we will be using a potential field
extrapolation employing a Fourier transform to solve for the
expansion of magnetic field into the half space above the
photosphere, as has been described already by 
e.g. \cite{Alissandrakis:1981}.

In a force free state, the Lorentz force has to vanish,
i.e. $\nabla{\times}\vec{B}$ has to be parallel to the magnetic field.
If one further assumes a current free state, then
\begin{equation} \nabla \times \vec{B} = 0 ~. \label{rotB=0} \end{equation} 
implying that $\vec{B}$ can be written in terms of a gradient of a
scalar field, i.e.\ $\vec{B}$ is a potential field.
This has to be accompanied, of course, by 
\begin{equation} \nabla \cdot \vec{B} = 0 ~.  \label{divB=0} \end{equation}



Starting from a given $z$-component of the 
magnetic vector in the $z{=}0$ plane and assuming a periodic nature 
of the problem in the horizontal directions, the field is calculated by utilizing
a Fourier transform (FT) technique to solve \eqn{rotB=0} and
\eqn{divB=0}.

\subsection{Expansion of the magnetic field} \label{expansion}

To solve the expansion of the magnetic field governed by \eqn{rotB=0}
and \eqn{divB=0} we follow \cite{Alissandrakis:1981}, but in
addition to his study we will also prescribe the ($z$-component of
the) magnetic field at the upper boundary.

The FTs, denoted by ${\cal F} \left[ ... \right]$, for the spatial
derivatives of $\vec{B}=(B_x,B_y,B_z)$ in the horizontal ($x$, $y$)
and vertical ($z$) directions read 
%
%
%
\begin{eqnarray}
{\cal F} \left[ \partial_x B_x \right] &=& ik_x \,{\cal F} \left[ B_x \right] \label{fdx}
\quad , \quad
{\cal F} \left[ \partial_y B_x \right] ~=~ ik_y \,{\cal F} \left[ B_x \right] \label{fdy}~, \nonumber\\
{\cal F} \left[ \partial_z B_x \right] &=& \partial_z {\cal F} [B_x] \label{fdz}~, \nonumber
\end{eqnarray}
and similar for $B_y$ and $B_z$. As the problem is assumed to be
periodic in $x$ and $y$, the wavenumbers $k_x$ and $k_y$ are real.
Applying these relations to the FTs of (\ref{rotB=0}) and
(\ref{divB=0}) yields a second order homogenous differential equation
\begin{equation} \partial^2_z \hat{B}_z = k^2 \hat{B}_z ~. \label{d2bz} \end{equation}
Here the horizontal wavenumber is given through
$k=(k^2_x+k^2_y)^{1/2}$ and for simplicity the hat symbol denotes the
Fourier transform, i.e. $\hat{B}_z = {\cal F} \left[ B_z \right]$.
For the horizontal components we find
%
\begin{equation}
\hat{B}_x = \frac{- i k_x}{k^2} \, \partial_z \hat{B}_z  \label{dbxy}
\qquad , \qquad
\hat{B}_y = \frac{- i k_y}{k^2} \, \partial_z \hat{B}_z ~.
\end{equation}
The general solution of (\ref{d2bz}) is
\begin{equation}
  \hat{B}_z(k_x{,}k_y{,}z) = a_1 \, e^{kz} + a_2 \, e^{- kz} ~, \label{ansatz}  \\
\end{equation}
with $a_1{=}a_1(k_x{,}k_y)$ and $a_2{=}a_2(k_x{,}k_y)$. 

When considering a computational domain stretching from zero to
infinity \citep[as done by][]{Alissandrakis:1981} in the 
vertical direction, naturally $a_1$ has to be zero! 
However, if we confine the considered
domain to a certain height we can specify a lower
\emph{and} an upper boundary condition (BC).
%
%
Based on the upper and lower BC we know the Fourier transforms of the
vertical component of the magnetic field at the bottom and the top of
our computational domain, 
\begin{eqnarray*}
 \hat{B}_{z_0} &=& \hat{B}_z(k_x,k_y,z{=}0) ~, \\ 
 \hat{B}_{z_h} &=& \hat{B}_z(k_x,k_y,z{=}h) ~.
\end{eqnarray*}
This determines the coefficients in the general solution (\ref{ansatz}),
$$
  a_1 = \frac{\hat{B}_{z_0} - \hat{B}_{z_h} \, e^{kh}}{1 - e^{2kh}}
  \qquad , \qquad
  a_2 = \frac{\hat{B}_{z_0} - \hat{B}_{z_h} \, e^{-kh}}{1 - e^{-2kh}} ~.
$$
All three components of the (FT of the) magnetic field can now be
written in terms of the assigned boundary conditions and height $z$,
\begin{eqnarray}
  \hat{B}_x &=&  \frac{-ik_x}{k} ~
    \frac{\hat{B}_{z_h} \sinh[kz] - \hat{B}_{z_0} \sinh[k(z{-}h)]}{\sinh[kh]} ~,
    \nonumber\\[1ex]
  \hat{B}_y &=&  \frac{-ik_y}{k} ~
    \frac{\hat{B}_{z_h} \sinh[kz] - \hat{B}_{z_0} \sinh[k(z{-}h)]}{\sinh[kh]} ~,
    \nonumber\\[1ex]
  \hat{B}_z &=&  
    \frac{\hat{B}_{z_h} \sinh[kz] - \hat{B}_{z_0} \sinh[k(z{-}h)]}{\sinh[kh]} ~.
     \nonumber
\end{eqnarray}
The magnetic field $\vec{B}$ is then determined by an inverse FT.

\subsection{Assigning Boundary conditions}

To achieve a maximum degree of comparability with the study of
\cite{Schrijver+Title:2003}, we employed magnetic configurations at
the lower boundary, $z{=}0$, which are derived by placing magnetic
charges below the surface as was done in their work. From these  we
extracted the $z$-component of the magnetic vector at the bottom
($z{=}0$) of  several computed field domains to create sets of
magnetograms for the use as lower BCs with the extrapolation model.

\subsubsection{Configuration of the lower boundary}\label{LBC}

In order to establish network-like magnetic patches we assume a
strong concentration of magnetic flux of $3{\cdot}10^{18}$\,Mx in the
middle of the horizontally periodic computational box, extending
14\,Mm $\times$ 14\,Mm horizontally.
This is then basically the same setup as in
\cite{Schrijver+Title:2003}, who considered 5$\times$5 magnetic
patches to achieve some periodicity.

Surrounding the central patch about 200 smaller charges of equal size
were randomly positioned to constitute the internetwork field.
Each of these small patches carries the same absolute value of
magnetic flux, but half of them has one, the other half the opposite
magnetic polarity, i.e.\ the net flux from the background, viz.\ the
internetwork, is zero.

We generated the magnetic field distribution at the lower boundary
by placing magnetic charges for the strong network patch as well as
for the smaller patches just below the surface 
(internetwork $z{=}{-}124$\,km \& network $z{=}{-}220$\,km)
and mirror charges at +100\,Mm.
Then we could compute the magnetic field by summing up the magnetic
potential at $z{=}0$.
In this way we ensured to have a lower boundary condition closely
matching the study of \cite{Schrijver+Title:2003}.
This results in a horizontal extension of the central network patch
at the photospheric level (full width at half maximum, FWHM) of
about 170\,km.
The surrounding weak internetwork patches have a FWHM of about
100\,km.
One can think of this lower BC for the magnetic field as a
magnetogram that should resemble a magnetic network patch surrounded
by a salt-and-pepper-like weak internetwork field.

We constructed several of such synthetic magnetograms to simulate
diverse flux distributions in the internetwork.
They all have the same central network patch carrying
$3{\cdot}10^{18}$\,Mx, but they differ with respect to the spatial
distribution and the magnetic flux of the weak internetwork patches.
We carried out experiments with the (absolute value of the) magnetic
flux for each weak internetwork patch ranging from 10$^{15}$\,Mx to
8$\cdot$10$^{17}$\,Mx.
Please note that in each of the synthetic magnetograms all
internetwork patches carry the same (absolute) flux.
The internetwork thus has an \emph{unsigned} mean flux density ranging from
0.1 to 80\,Mx/cm$^2$ (while the total flux is zero).
The overall net flux density is always 1.5 Mx/cm$^2$,  since only the
central magnetic concentration contributes to the net flux.
(See \tab{in_conf}).

\begin{table}
\begin{center}
\caption{%
Parameters of the lower boundary condition, i.e.\ for the
strong network patch and the 200 surrounding smaller internetwork
flux concentrations (see \sect{LBC}).
$\phi_i$ denotes the magnetic flux carried by each individual patch,
$A$ is the horizontal extent of the bottom boundary.
\label{in_conf}}
\begin{tabular}{l@{~}l@{}cc} \hline
                           && network            
                            & internetwork 
\\ \hline
\multicolumn{2}{l@{}}{size of magnetic patches}\\
{\sc fwhm} & [km]              & 169                
                               &  96      \\[0.8ex]
\multicolumn{2}{l@{}}{flux of magnetic patches}\\
$|\phi_i|$ & [Mx]              & $3{\cdot}10^{18}$  
                               & $10^{15} \ldots 8{\cdot}10^{17}$  \\[0.8ex]
\multicolumn{2}{l@{}}{unsigned total flux}\\
$\sum|\phi_i|$ & [Mx]          & $3{\cdot}10^{18}$ 
                               & $2{\cdot}10^{17} \ldots 1.6{\cdot}10^{20}$  \\[0.8ex]
\multicolumn{2}{l@{}}{unsigned mean flux density}\\
$\sum|\phi_i|/A$ & [Mx/cm$^2$] & 1.5 & $0.1 \ldots 80$  \\[0.8ex]
\multicolumn{2}{l@{}}{net flux} \\
$\sum \phi_i$ & [Mx]           & $3{\cdot}10^{18}$ & 0 \\[0.8ex]

\hline
\end{tabular}
\end{center}
\end{table}

Concerning the spatial distribution we constructed 22 sets of randomly
distributed internetwork patches.
We then calculated a magnetogram for each of the 22 spatial
distributions with each of the 19 sets of internetwork fluxes.
Together with the 4 conditions discussed in the next subsection this
adds up to a total of 1672 configurations.

Through this we can study independently (1) the impact of the
strength of the magnetic flux in the internetwork and (2) the role of
the spatial distribution of magnetic patches in the internetwork,
both at the bottom boundary, i.e.\ the photospheric level.

\subsubsection{Configuration of the upper boundary}\label{UBC}

In contrast to the work of \cite{Schrijver+Title:2003} we will also
investigate the role of the upper boundary on the flux budget from
the network patches into the corona.
To explore this we employed four different upper BCs for the
magnetic field extrapolation outlined in \sect{expansion}.
These are partly carried to the extreme (especially item 4) and are
not necessarily realistic, but have been chosen to illustrate
possible effects of the upper boundary.

{(1)}
In the first case we assume the field to be stretching to infinity
\citep[in the same way as e.g.][]{Alissandrakis:1981}, i.e.\
$a_1{=}0$ in (\ref{ansatz}).

{(2)}
The next case uses a $B_z(x{,}y)$ configuration at $z{=}15$\,Mm
extracted from the extrapolation method used also by
\cite{Schrijver+Title:2003} as an upper BC.
By this we match the $B_z$-component at $z{=}15$\,Mm with their work
for a reliable comparison.

{(3)} 
In a third case we applied a 2D Gaussian curve on top of a constant
field at a height of $z{=}15$\,Mm at the upper boundary, i.e.\ a 
broad (FWHM 6\,Mm wide) maximum just above the central
network patch.
In order to have a realistic $B_z$ variation at the upper boundary,
we examined an actually observed magnetogram (obtained
with MDI) for a region that showed a magnetic configuration similar
to what we have considered the quiet Sun in this work.
Applying a standard potential field extrapolation, with an upper
boundary as for case (1), we derived a $B_z$ variation of some
$\approx$5.5\% at $z{=}15$\,Mm.
This we used for the amplitude of the Gaussian above the background.

{(4)}
For the last case we deployed a very unrealistic  magnetic
configuration at $z{=}15$\,Mm to see what effect a really extreme upper
boundary might have.
For this we used the same Gaussian as in case (3), but now with $-$50\%
variation, i.e.\ with the magnetic field strength being depressed by
a factor of 2 at the upper boundary directly above the central
network patch.

\subsection{Budget of magnetic flux}

Once three-dimensional grids of complete $\vec{B}(x,y,z)$ sets were
computed through the extrapolation as described above we determined
field lines by  tracing paths of monopolar test particles through
the box.

\subsubsection{Field line tracing}

\cite{Schrijver+Title:2003} traced field lines starting in random 
directions from the central network patch in the $z{=}0$
plane. Their method then follows magnetic field lines until they
either connect back onto the source plane or reach a maximum
height, i.e.\ an upper boundary.

One of the main goals of our work is to find the fraction of magnetic 
flux from the strong network patch, that has made its way into the
corona, i.e.\ to the upper boundary.
Therefore we start our field line tracing at the top boundary
($z{=}15$\,Mm).
Once lines reach the $z{=}0$ plane, we count them either as connecting
the network or the internetwork to the corona, depending whether or
not they are coming down on the central network patch.

We start tracing field lines from a regular grid at the upper
boundary with a spacing of 100\,km. 
By this reverse approach we achieve a high precision when answering
the question where the source of coronal magnetic flux is located.

\begin{figure}
\begin{center}
\includegraphics[width=\figwid, bb=226 340 694 848,clip=true]{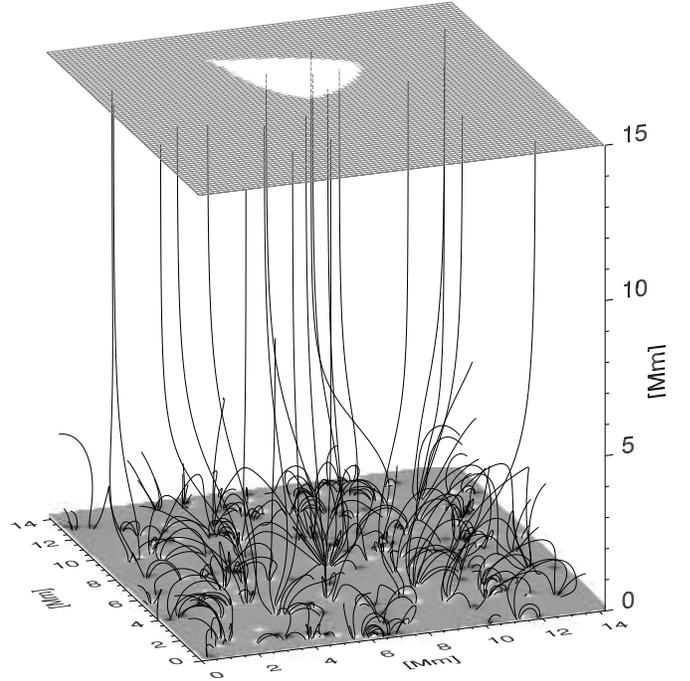}
\caption{Composite plot of the magnetic field at the bottom (lower
  BC), field lines of the extrapolated magnetic field, and a mask of
  the area at 15\,Mm height that is magnetically connected to the
  central strong network magnetic patch (white).
  \label{deckel}}
\end{center}
\end{figure}

For the purpose of plotting the filed line configurations as shown in
\fig{deckel} and \ref{coupling_hoehe} we trace the field lines also
from the bottom boundary, of course.
\fig{deckel} shows the field lines for a sample potential field
extrapolation.
At the bottom of the figure the distribution of the (vertical) magnetic
field is displayed, while the upper part of the plot shows a mask,
where the area at a height of 15\,Mm connected to the central strong
network magnetic field patch is plotted in white, while the area
connected to the weak surrounding internetwork patches is shaded
gray.

\subsubsection{Connecting network and corona} \label{connecting}

The equivalent flux carried by each field line counts either towards
the network or the internetwork flux contingent.
The amount of magnetic flux associated with a given field line is
calculated through
\[ \phi_i =  B_z(x_i{,}y_i{,}z{=}h) \cdot A_i ~,  \]
where $A_i$ denotes the area of the resolution element at the top
boundary to which the field line is connected to, i.e.\ the grid spacing
of 100\,km ${\times}$ 100\,km. 
Respectively $B_z(x_i{,}y_i{,}z{=}h)$  is the field strength at that
resolution element.

The sum of those fluxes $\phi_i$ associated with the field lines connecting the network
patch to the upper boundary gives the flux $\Phi^{Cor}_{Netw}$ in the corona
originating from the magnetic network element. Comparing this to the total
magnetic flux of the network patch $\Phi^{tot}_{Netw}$, i.e. 
\begin{equation}\label{fraction}
f=\Phi^{Cor}_{Netw} / \Phi^{tot}_{Netw} ~,
\end{equation}
yields the fraction of magnetic flux from the network which is connected to the 
corona.

\section{Results and Discussion}

We applied the extrapolation method and the calculation of the flux
budget from the network patch into the corona for all of the 1672
sets of boundary conditions, viz.\ photospheric magnetograms.
As a subset these also include the approach used by
\cite{Schrijver+Title:2003}.

The fraction of magnetic flux that expands from the network patch to
reach the corona  was calculated for four distinct
$B_z$-configurations at the upper boundary (cf.\ \sect{UBC}) combined
with 22 different (random) distributions of magnetic patches in the
internetwork field (cf.\ \sect{LBC}).
Each of these 88 configurations was repeated for 19 different
unsigned mean magnetic flux densities of the  internetwork field
ranging from 0.1 to 80 Mx/cm$^2$ (cf.\ \sect{LBC}).

In the following we compare our results with those presented by
\cite{Schrijver+Title:2003} in \sect{comparison},  investigate the impact
of the upper boundary (\sect{impact_ubc}) and study the role of the
lower boundary (\sect{impact_lbc}).

\subsection{Comparison to previous work}\label{comparison}

\begin{figure}
\begin{center}
\includegraphics[width=\figwid]{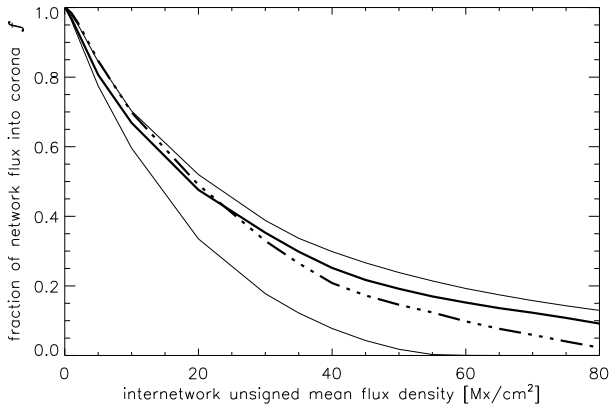}
\caption{%
Average fraction $f$ of magnetic flux from the strong network patch
into the corona as defined in \eqn{fraction} as a function of the
unsigned mean internetwork flux density.
The dot-dashed line shows the result for a point charge model similar
to \cite{Schrijver+Title:2003}, see \sect{comparison}.
The thick solid line displays $f$ for the field extrapolation
using a Fourier transform technique.
For all (!) of the four upper boundary conditions discussed in
\sect{UBC} the results are practically the same.
While the thick line represents the median value of $f$ for the
different spatial random distributions of internetwork patches,
the thin lines show the scatter (enclosing 2/3 of all values for
$f$).
See \sect{comparison} and \ref{impact_ubc}.
\label{schr_wir}}
\end{center}
\end{figure}

Recalculating the models of \cite{Schrijver+Title:2003}
including an evaluation of the flux budget enabled us to positively 
verify their results.
As they perform the extrapolation using a distribution of magnetic
charges, their study  might be called ``point charge model''.
Based on such a model we calculated the average value of the
fraction $f$ defined in \eqn{fraction} of magnetic flux from the
network patch being connected to the corona for the various
distributions of internetwork flux at the lower boundary (\sect{LBC}).
The dot-dashed line in \fig{schr_wir} shows the average fraction $f$
as a function of the unsigned mean internetwork flux density.
We find the same roughly exponential decrease of the fraction $f$ as
\cite{Schrijver+Title:2003}, differing by 0.05 to 0.08 as compared to
their results.
This deviation basically stems from using either field line tracing
starting from above (\sect{connecting}) as done in this work or the
method of \cite{Schrijver+Title:2003} starting from below.

We can now also compare the point charge model with a potential field
extrapolation utilizing a Fourier transformation (FT) technique as
outlined in \sect{expansion} with a upper BC as given in case 2 of
\sect{UBC}, i.e.\ which has the same magnetic field distribution at
the upper boundary as the point charge model.
The result for the average fraction $f$ for this experiment is plotted as a
thick solid line in \fig{schr_wir}.
The thin solid lines indicate the scatter of the $f$ values derived
for the different spatial distributions of the weak internetwork flux
patches.
The difference of the point charge model (dot-dashed) and the FT
technique (solid) is rather small, which shows that the two methods
give similar results.
Thus we regard it as justified that both methods are likewise
suitable  to tackle the problem at hand. In our work we proceed using
the extrapolation model  in order to be able to determine the possible
impact of the upper boundary, which could not be handled by the point
charge model.

\subsection{Impact of the lower corona on the network coupling}
                                             \label{impact_ubc}

We first turn to the impact of the coronal configuration on the
connectivity from the chromosphere into the corona, i.e.\ the role of
the upper boundary.
Depending on the large-scale structures in the corona one might
imagine different magnetic configurations at the low corona.
For example, while one might expect a rather constant magnetic field
at a height of some 15\,Mm in a more or less unipolar coronal hole,
one could expect a rather strong variation in a network patch located
near an active region.
To explore this we employ four different sets of upper BCs (see
\sect{UBC}).

For these cases we evaluated the fraction $f$ of magnetic flux from the
network element reaching the corona in the same way as in the
previous section.
We find that for all~(!) cases of the upper BC the resulting
fractions $f$ are practically the same.
In other words, comparing the budget of magnetic flux connecting the
network patch and the corona for substantially diverse
$B_z$-configurations at the upper boundary, we find no
evidence of an impact of the magnetic field distribution in the low
corona on the magnetic connection from the chromosphere to the corona.
This is true even for the quite ``pathological'' case 4 outlined in
\sect{UBC}, with a very strong (unrealistic) depression of the
magnetic field above the network patch by a factor of 2.

\begin{figure}
\begin{center}
\includegraphics[width=\figwid, bb=150 210 510 504,clip=true]{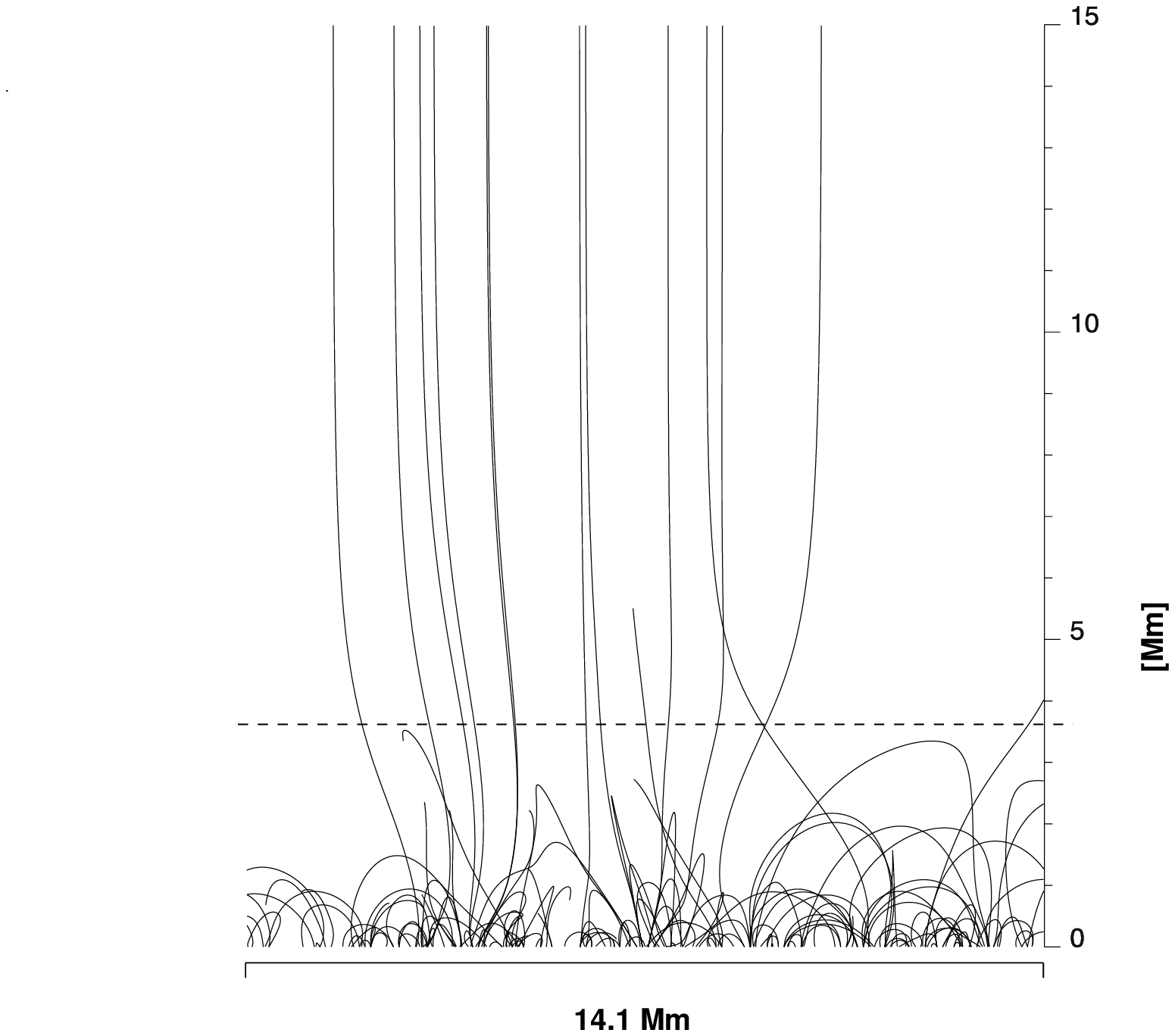}
\caption{%
A selection of field lines starting from  the internetwork are
drawn for an unsigned mean internetwork flux density of 60
Mx/cm$^2$. Here the computational domain is viewed from the side.
The dashed line indicates the maximum height of the
highest-reaching closed field line from this selection. See \sect{impact_ubc}.
\label{coupling_hoehe}}
\end{center}
\end{figure}

This result can be illustrated by a display of magnetic field lines
when looking at the computational domain from the side (see
\fig{coupling_hoehe}).
Above an altitude of about 5\,Mm the magnetic field is basically
open, i.e. all the field lines found above that magnetic transition
are connected to the corona (the dashed line in \fig{coupling_hoehe}
shows the height of the highest-reaching closed field line in that selection).
Above this magnetic transition the small scale ``salt and pepper''
structure of the internetwork magnetic field is cleared and a more or
less vertical $B$-component dominates.
The height of that magnetic transition is, of course, related to the
mean distance of the magnetic polarities in the photosphere, which is
for the distributions of the present work as well as on the real Sun of
the order of a couple of Mm.

The good news of this little experiment is, that any physically
reasonable assumption for the  magnetic field at the upper boundary is
sufficient, supposing it is assigned at an altitude above the magnetic
transition, i.e.\ above some 5\,Mm.
This is of special interest, as we are not really able to measure
coronal magnetic fields as of yet, except for some special cases
\citep[e.g.][]{Lin+al:2000,Lagg+al:2004,White:2005}.
Thus even without a proper information on the magnetic field in the
low corona we can study the magnetic connectivity from the
chromosphere into the corona by means of magnetic field
extrapolations in a meaningful way on a granular and super-granular
scale.

\subsection{Impact of the photospheric field distribution on the magnetic connectivity}
                                        \label{impact_lbc}

We will now turn to the role of the lower boundary, i.e.\ the
photospheric magnetic field for the magnetic connectivity into the
corona.
As emphasized by \cite{Schrijver+Title:2003} a sizable fraction
of the corona is connected not to the strong network patches, but
to the weak internetwork fields.
Thus when trying to relate coronal phenomena to chromospheric events,
one cannot simply look for e.g.\ a correlation of the corona with a
network patch below, but coronal events can also be triggered by
processes in the internetwork.
The basic question is here if we can use the magnetic field
extrapolations to derive which coronal structures are connected to
which areas in the photosphere and chromosphere, no matter if in the
network or internetwork.

To tackle this question we will investigate the spatial distribution
of the weak internetwork magnetic field patches in our simulated
magnetograms, which are hardly resolved by current instrumentation.
For example MDI onboard SOHO \citep{Scherrer+al:1995:mdi} has a
sensitivity of some 20 Gauss (if being optimistic) and a spatial
resolution in the high-resolution modus of 1\,arcsec
($\approx$0.5\,arcsec pixels) corresponding to 725\,km.
Thus this instrument, which is often used to study the relations of
magnetic field and coronal phenomena, can detect magnetic flux
concentrations down to some $10^{17}$\,Mx.
Therefore most of the small internetwork flux patches ranging from
$10^{15}$ to $8{\cdot}10^{17}$\,Mx as used in this study (cf.\
\sect{LBC}) are not detectable with MDI.
And even for todays most sophisticated high-resolution
spectro-polarimeters this is a difficult task, especially at the low
flux end.
For example {\sc Polis} at the German VTT in Tenerife \citep{Beck+al:2005} can
go down to some 5 Gauss with a resolution of 0.5\,arcsec, if seeing
conditions permit, i.e.\ down to fluxes of some $10^{16}$\,Mx.

\begin{figure*}[!t]
\hbox{
\includegraphics[height=\figwff]{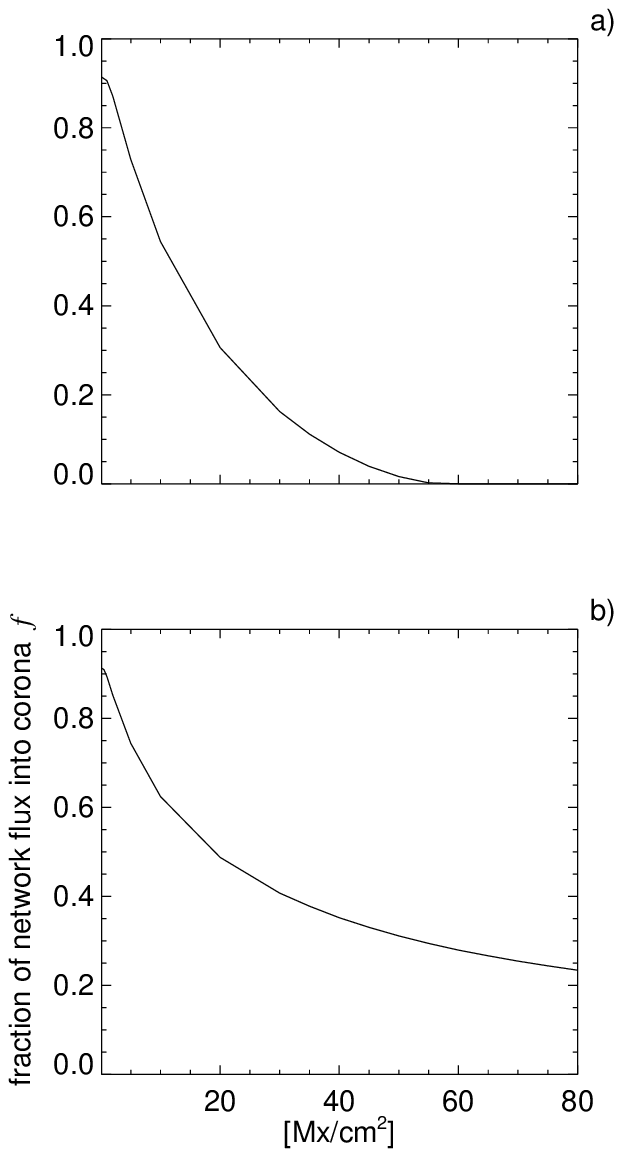}      
\includegraphics[height=\figwff]{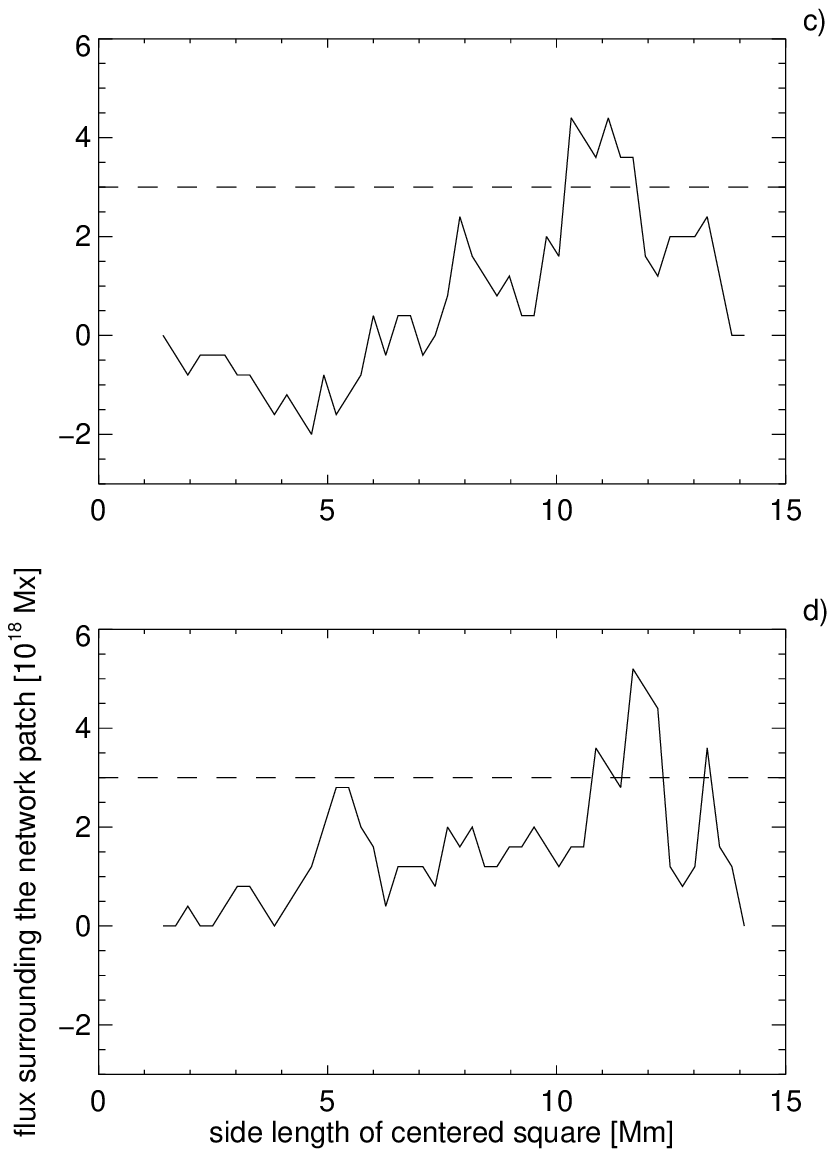}      
\includegraphics[height=\figwff]{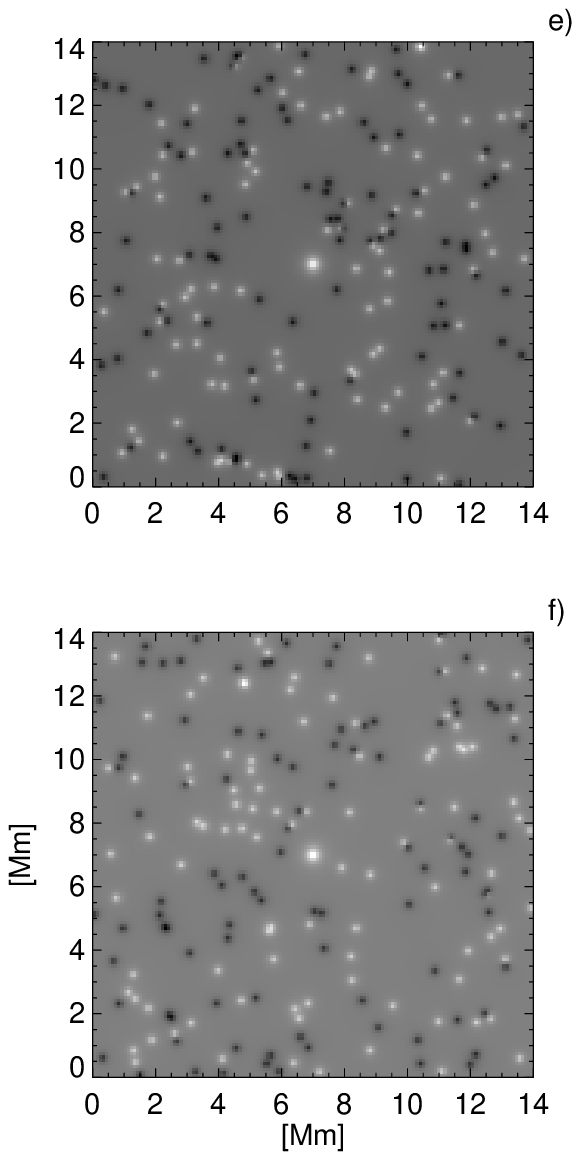}}      
\caption{%
Individual examples of spatial distributions of weak internetwork
patches. Each row shows the results for one spatial sample distribution.
\emph{Right colum:} 
Lower boundary condition for the vertical component of the magnetic
field shown as a magnetogram with black and white representing
opposite polarities. The strong network patch is visible right in the
middle of the magnetogram.
\emph{Middle column:}
Amount of internetwork flux contained in a square with length of a side
$a$ centered around the central network patch for lengths $a$
ranging from outside the central network patch to the full coverage
of the horizontal extent of the computational box. 
Internetwork unsigned mean flux density: 40 Mx/cm$^2$. 
The dashed line marks the magnetic flux of $3{\cdot}10^{18}$\,Mx that 
emanates from the network element which in \emph{d} is almost 
balanced by the internetwork in a short distance of roughly
2.5 Mm from the 
center.
\emph{Left column:}
Fraction $f$ of magnetic flux from the strong network patch
into the corona as defined in \eqn{fraction} as a function of the
unsigned mean internetwork flux density.
Please note the significant difference in network flux connected to
the corona for the examples shown here in the top and bottom row.
See \sect{res_spatial}.
\label{gesamt}}
\end{figure*}

In conclusion, especially the small internetwork flux concentrations
we inverstigatetd here are hard to resolve; with some $10^{16}$\,Mx
they correspond to a 1000\,G flux tube with 36\,km diameter.
As there is no real reason to belive that such small structures do
not exist, e.g.\ they are found in numerical simulation of
magneto-convection \citep[e.g.][]{Voegler+al:2005}, it is an
interesting task to investigate the influence of such small
salt-and-pepper-like flux concentrations to the magnetic
connectivity into the corona.

\subsubsection{Connectivity and flux distribution}\label{res_spatial}

To investigate the role of the actual spatial distribution of
magnetic flux at the bottom, viz.\ the photosphere, it is instructive
to study individual examples of the random distributions of the
internetwork flux patches.

Two examples are shown in \fig{gesamt}. 
The right column shows the lower boundary conditions for both cases,
i.e. the magnetograms constructed from the random distributions of
flux patches of the internetwork and the network patch in the middle.
The images display the vertical component of the magnetic field on
a grey scale, with black and white representing opposite polarities.
As outlined in \sect{connecting} we calculated the fraction $f$ of
flux from the network patch reaching the corona for various values of
the flux of the internetwork patches (but respectively always for the
same spatial distribution).

The resulting fraction $f$ as a function of unsigned mean
internetwork flux density is shown in the respective graphs in the
left column of \fig{gesamt}.
As can be easily seen the fraction of network flux reaching the
corona is quite different in the two sample spatial distribution.
While in one case (top row in \fig{gesamt}) basically all coronal flux
is originating from the internetwork (!) for average unsigned
internetwork fluxes above some 50\,Mx/cm$^2$, in the other case
always more than some 20\% of the network flux reach the corona, even
for very strong internetwork fields (bottom row).
On average, i.e.\ when considering a large number of random spatial
distributions of the internetwork patches, one ends up with a
fraction $f$ somewhere between the two extremes shown in
\fig{gesamt}, and this average result was already discussed in
\sect{comparison} and \fig{schr_wir}.
Of course, this is simply due to the distribution of flux directly
surrounding the central strong network patch.
The middle column of \fig{gesamt} shows the total internetwork flux
surrounding the network patch in a square with a length of
a side $a$, i.e.\ excluding the network patch.
In one case (top row) the surroundings of the central (positive)
network patch are dominated by negative charges, in the other case
(bottom row) the opposite is true.
Thus depending on the dominating sign of magnetic flux near the
network patch one will get a very low or a very high fraction of the
network flux being connected to the corona.

In a way, there is a chance for special configurations of the
internetwork flux patches to ``neutralize'' the strong network patch,
so that most of the coronal magnetic field actually originates from
the internetwork.

The intriguing part of this (in principle obvious) result is as
follows.
If the internetwork field patches are hardly resolvable by
observations, but still can significantly alter the amount of flux
from the network reaching the corona, we are confronted with a clear
limitation for current investigations of the interaction between the
chromosphere and the corona.

\subsubsection{Mapping the corona back to the photosphere}

So far we have examined the connectivity between the photospheric
field and the corona only quantitatively through the fraction $f$ of
network flux reaching the corona.
We will now turn to mapping coronal regions back to the photosphere
and vice versa.

\begin{figure}
\includegraphics[height=\figwid]{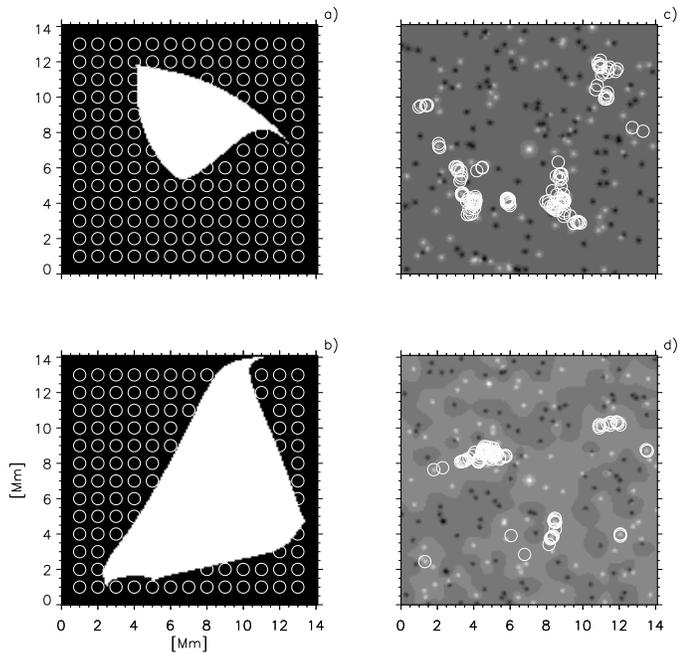}
%
\caption{Mapping from the photosphere to the corona.
The left panels show in white the areas in the corona at an altitude
of 15\,Mm which are connected to the strong central network patch for
two distributions of internetwork magnetic patches (same as the
examples shown in \fig{gesamt}).
The black regions are connected to the internetwork.
The right panels show maps of the vertical magnetic field at the
bottom boundary, i.e.\ in the photosphere (re-plotted from
\fig{gesamt}).
Over-plotted are white circles which indicate the intersections of
the magnetic field lines connecting the internetwork patches with the
corona at the upper and lower boundaries, i.e. the circles in the
right panles show the footpoints of the fieldlines connecting weak
internetwork patches with the corona.
\label{blackwhite}}
\end{figure}

The masks in the left panels of \fig{blackwhite} show in white the
areas in the corona at a height of 15\,Mm which are connected to the
central network patch in the photosphere, i.e.\ at the lower boundary.
Here we used the same sample distributions of weak internetwork
magnetic flux patches as in the previous subsection and \fig{gesamt}.

This reveals that the corona is quite sensitive to the structure of
the weak internetwork fields: the coronal volume which is connected
to the strong magnetic network patches changes significantly, if a
different distribution of weak internetwork field patches is
prescribed.
Our computations show that this is valid even down to
very low flux concentrations ($0.1$--$2{\cdot}10^{16}$\,Mx) in the
internetwork, which are  hardly resolvable at the current state of
observational capabilities.

It is not only problematic to derive the coronal area(s) connected to
the network patch, but also to identify which parts of the internetwork
are connected to the corona seems hardly possible when using current
instrumentation.
This is illustrated also by \fig{blackwhite}, where the intersections
with the top and bottom boundary of the fieldlines connecting the
internetwork with the corona are indicated by circles.
From the right panels it is evident, that the internetwork footpoints
of the fieldlines connecting to the corona cluster at completely
different regions, depending on the distribution of the photospheric
internetwork flux.

These results have serious implications for studies of the
relation of the photospheric field to the coronal structures, e.g.\ 
as been done in the context of solar wind acceleration and its
relation to the chromospheric magnetic network
\cite[e.g.][]{Xia+al:2004,Tu+al:2005}.
When using data from the MDI instrument, as mostly done in studies
relating the corona to the photosphere in recent years, one is not
able to resolve the weak internetwork fields as modeled in the
present paper. 
This implies, that in those studies one cannot definitely pin down
which areas in the low corona are in fact connected to which parts of
the photosphere.
Certainly, under the presence of weak fields, the connection from the
coronal features down to the photosphere will not be nice and
funnel-type, but more spaghetti-type with a lot of connections from
the coronal patches to the internetwork.
Better future instrumentation with superior sensitivity for measuring
photospheric fields will hopefully provide the necessary
information to allow advanced studies of the connectivity from the
photosphere to the corona.


\section{Conclusions}		\label{conclusions}

In the present paper we investigated the role of the magnetic field
distribution in the photosphere and the low corona for the
mapping from the chromospheric magnetic network into the corona
utilzing magnetic field extrapolations.
The classic picture was that the field fans out from strong network
concentrations  in the chromosphere creating an often called canopy
\citep{Gabriel:1976}, which could be seen as a more or less stable
foundation of the coronal magnetism.
Extending the work of \cite{Schrijver+Title:2003}, who are
challenging this scenario, we manipulate the boundary conditions of a
potential field extrapolation at the top \emph{and} bottom of a
computational box stretching from the photosphere to the low corona
at some 15\,Mm height.
This allows us to investigate the mapping from the photosphere into
the corona in great detail.

The good news are that the photosphere--corona connectivity is very
stable against changes of the top boundary, i.e.\ the magnetic field
in the lower boundary at heights of around 15 Mm.
This is very reassuring when studying different regions on the Sun,
such as coronal holes, active region patches or quiet Sun, where one
might expect different magnetic field structures in the low corona:
when investigating the connection e.g.\ of coronal Doppler shifts and
photospheric fields to study the driving of coronal dynamics, a
magnetic field extrapolation from an observed high-quality
magnetogram seems to be sufficient to identify the respective areas
in the photosphere and the corona which are magnetically connected
(cf.\ \sect{impact_ubc}).

However, the reliability of the magnetic connection derived from the
field extrapolation hinges on the quality of the information about
the weak small-scale magnetic field in the photosphere.
Changes in the distribution of the weak internetwork magnetic flux
patches, which are hardly observable with current instrumentation,
can lead to dramatic differences in the connectivity from the
photosphere into the chromosphere, qualitatively as well as
quantitatively.
Especially the mapping from the photosphere into the corona, i.e.\
which photospheric regions are magnetically connected to which
coronal areas, is quite sensitive to the distribution of weak
internetwork flux.
With current instrumentation only beginning to resolve the structure
of the internetwork magnetic field, one has to be careful when
drawing conclusions on magnetic connections from network patches to
the corona, as one cannot really tell about the role of the
internetwork (cf.\ \sect{impact_lbc}).

It will be exciting to investigate the polarimetric studies with new
high-resolution instruments from space as well as from the ground, e.g.\
the soon to be started Solar\,B mission or the 1.5\,m solar telescope
{\sc{Gregor}} with their sensitive full Stokes polarimeters.
Such studies will be of great importance to better understand the
driving of the corona through the underlying photospheric convection.





\def\apj       {ApJ}%
\def\apjl      {ApJ}%
\def\apjs      {ApJS}%
\def\apss      {Astrophys. Space Sci.}%
\def\jgr       {J. Geophys. Res.}%
\def\nat       {Nature}%
\def\solphys   {Solar Phys.}%
\def\philtrans {Phil.\ Trans.\ Roy.\ Soc.\ Lond.}%
\def\aap       {A\&A}%
\def\aaps      {A\&A Suppl.}%
\def\mnras     {Mon.\ Not.\ Roy.\ Astron.\ Soc.}
\def\araa      {Ann.\ Rev.\ Astron.\ Astrophys.}
\def\naturwiss {Naturwiss.}
\def\za        {Zeitschrift f.\ Astrophys.}
\def\an        {Astronomische Nachrichten}
\def\pasp      {PASP}

{\small


}

\end{document}